\documentclass[useAMS]{mn2e}

\usepackage{xspace}
\include{epsf}
\newcommand{\hi}{H\,{\sc i}}

\newcommand{\hicat}{{\sc Hicat}\xspace}
\newcommand{\hipass}{{\sc Hipass}\xspace}

\newcommand{\kms}{\mbox{$\rm km\, s^{-1}$}}
\newcommand{\jykms}{\mbox{$\rm Jy\, km\, s^{-1}$}}
\newcommand{\mhi}{\mbox{$M_{\rm HI}$}}
\newcommand{\speak}{\mbox{$S_{\rm p}$}\xspace}
\newcommand{\speakf}{S_{\rm p}}
\newcommand{\sint}{\mbox{$S_{\rm int}$}}

\newcommand\mnras{{MNRAS}}%
\newcommand\aap{{A\&A}}%
\newcommand\aaps{{A\&AS}}%
\newcommand\apj{{ApJ}}%
\newcommand\apjs{{ApJS}}%
\newcommand\aj{{AJ}}%
\title[The HIPASS Catalogue -- II]{The HIPASS catalogue -- II. Completeness,
  reliability, and parameter accuracy}

\author[M.A. Zwaan et al.]
{M. A. Zwaan,$^{1,2}$\thanks{E-mail: mzwaan@eso.org (MZ), martinm@stsci.edu (MM), rwebster@ph.unimelb.edu.au (RW), lister.staveley-smith@csiro.au (LSS) }
M. J. Meyer,$^{1,3}$
R. L. Webster,$^{1}$
L. Staveley-Smith,$^{4}$\newauthor
M. J. Drinkwater,$^{5}$
D. G. Barnes,$^{1}$
R. Bhathal,$^{6}$
W. J. G. de Blok,$^{7}$\newauthor
M. J. Disney,$^{7}$
R. D. Ekers,$^{4}$
K. C. Freeman,$^{8}$
D. A. Garcia,$^{7}$
B. K. Gibson,$^{9}$\newauthor
J. Harnett,$^{10}$
P. A. Henning,$^{11}$
M. Howlett,$^{9}$
H. Jerjen,$^{8}$
M. J. Kesteven,$^{4}$\newauthor
V. A. Kilborn,$^{12,9}$
P. M. Knezek,$^{13}$
B. S. Koribalski,$^{4}$ 
S. Mader,$^{4}$
M. Marquarding,$^{4}$\newauthor
R. F. Minchin,$^{7}$
J. O'Brien,$^{8}$ 
T. Oosterloo,$^{14}$
M. J. Pierce,$^{9}$
R. M. Price,$^{11}$\newauthor 
M. E. Putman,$^{15}$
E. Ryan-Weber,$^{1,4}$
S. D. Ryder,$^{16}$
E. M. Sadler,$^{17}$ 
J. Stevens,$^{1}$\newauthor
I. M. Stewart,$^{18}$
F. Stootman,$^{6}$   
M. Waugh,$^{1}$
and A. E. Wright$^{4}$\\
$^{1}$ School of Physics, University of Melbourne, VIC 3010, Australia.\\
$^{2}$ European Southern Observatory, Karl-Schwarzschild-Str. 2, 85748 Garching b. M{\"u}nchen, Germany.\\
$^{3}$ Space Telescope Science Institute, 3700 San Martin Drive, Baltimore MD 21218, USA.\\
$^{4}$ Australia Telescope National Facility, CSIRO, P.O. Box 76, Epping, NSW 1710, Australia.\\
$^{5}$ Department of Physics, University of Queensland, QLD 4072, Australia.\\
$^{6}$ Department of Physics, University of Western Sydney Macarthur, P.O. Box 555, Campbelltown, NSW~2560, Australia.\\
$^{7}$ Department of Physics \& Astronomy, University of Wales, Cardiff, P.O. Box 913, Cardiff CF2 3YB, U.K.\\
$^{8}$ Research School of Astronomy \& Astrophysics, Mount Stromlo Observatory, Cotter Road, Weston, ACT~2611, Australia.\\
$^{9}$ Centre for Astrophysics and Supercomputing, Swinburne University of Technology, P.O. Box 218, Hawthorn, VIC 3122 Australia.\\
$^{10}$ University of Technology Sydney, Broadway NSW 2007, Australia.\\
$^{11}$ Institute for Astrophysics, University of New Mexico, 800 Yale Blvd, NE, Albuquerque, NM~87131, USA.\\
$^{12}$ Jodrell Bank Observatory, University of Manchester, Macclesfield, Cheshire, SK11 9DL, U.K.  \\
$^{13}$ WIYN, Inc. 950 North Cherry Avenue, Tucson, AZ, USA.\\
$^{14}$ ASTRON, P.O. Box 2, 7990 AA Dwingeloo, The Netherlands.\\
$^{15}$ CASA, University of Colorado, Boulder, CO 80309-0389, USA.\\
$^{16}$ Anglo-Australian Observatory, P.O. Box 296, Epping, NSW~1710, Australia.\\
$^{17}$ School of Physics, University of Sydney,  NSW~2006, Australia.\\
$^{18}$ Department of Physics \& Astronomy,  University of Leicester, Leicester LE1 7RH, U.K.}

\begin{document}

\date{Accepted ...
      Received ...}

\pagerange{\pageref{firstpage}--\pageref{lastpage}}
\pubyear{0000}

\maketitle

\label{firstpage}

\begin{abstract}
The \hi\ Parkes All Sky Survey (\hipass) is a blind extragalactic \hi\
21-cm emission line survey covering the whole southern sky from
declination $-90^\circ$ to $+25^\circ$. The \hipass catalogue
(\hicat), containing 4315 \hi-selected galaxies from the region south
of declination $+2^\circ$, is presented in Meyer et al. (2004a, Paper
I). This paper describes in detail the completeness and reliability of
\hicat, which are calculated from the recovery rate of synthetic
sources and follow-up observations, respectively.  \hicat is found to
be 99 per cent complete at a peak flux of 84~mJy and an integrated
flux of $9.4~\jykms$.  The overall reliability is 95 per cent, but
rises to 99 per cent for sources with peak fluxes $>58~\rm mJy$ or
integrated flux $> 8.2~\jykms$.  Expressions are derived for the
uncertainties on the most important \hicat parameters: peak flux,
integrated flux, velocity width, and recessional velocity. The errors
on \hicat parameters are dominated by the noise in the \hipass data,
rather than by the parametrization procedure.
\end{abstract}
 

\section{Introduction}
The \hi\ Parkes All Sky Survey (\hipass) is a blind 
neutral hydrogen survey over the entire sky south of
declination $+25^\circ$. One of the main objectives of the survey is
to extract a sample of \hi-selected extragalactic objects, which can
be employed to study the local large scale structure and the
properties of galaxies in a manner free from optical selection
effects.  In Meyer et al. (2004a, paper I, hereafter) we present the
\hipass sample of 4315 \hi-selected objects from the region south of
declination $+2^\circ$.  This sample, which we refer to as \hicat,
forms the largest catalogue of extragalactic \hi-selected objects to
date.  In paper I the selection procedure of \hicat is described in
detail, along with a discussion of the global sample properties 
and a description of the catalogue parameters that have been released
to the public. The scientific potential of \hicat is very large, but to
make optimal use of the catalogue it is essential that the
completeness and reliability are well understood and quantified.  Only
after an accurate assessment of the completeness and reliability is it
possible to extract from the observed sample the intrinsic properties
of the local galaxy population.

For optically-selected galaxy samples, this procedure is relatively
straightforward since most optically-selected galaxy samples are
purely flux-limited, possibly complicated by the reduced detection
efficiency of objects with low optical surface brightness (see e.g.,
Lin et al 1999, Strauss et al. 2002, Norberg et al. 2002).  Since the
\hi\ 21-cm emission of galaxies is localized in a narrow region of
velocity space, blind 21-cm surveys need to cover the two spatial
dimensions and the velocity dimension simultaneously. The advantage of
this is that the survey yields redshifts simultaneously with the
object detections, and follow-up redshift surveys are not
required. However, this extra dimension complicates the detection
efficiency. The `detectability' of a 21-cm signal depends not
only on the flux, but also on how this flux is distributed over the
velocity width of the signal.

In this paper we take an empirical approach to this problem, and
determine the completeness of \hicat by the recovery rate of synthetic
sources that have been inserted in the data. The reliability is
determined by follow-up observations of a large number of sources.
Our aim is to describe in detail the completeness and the reliability
of \hicat as a function of various catalogue parameters, in such a way
that future users can make optimal use of \hicat in studies of e.g.,
the \hi\ mass function, the local large scale structure, the
Tully--Fisher relation, etc.  We also discuss in detail the errors on
the \hicat parameters, determine expressions to estimate errors and
estimate what fraction of the error is determined by noise and what
fraction by the parametrization.

The organization of this paper is as follows: in
Section~\ref{hipass.sec} a brief review of the \hipass surveys is
given. In Section~\ref{completeness.sec} the completeness of \hicat is
calculated using three independent
methods. Section~\ref{reliability.sec} details the follow-up
observations and the evaluation of the reliability. In
Section~\ref{parameter.sec} errors on \hicat parameters are calculated.

\section{The HIPASS survey} \label{hipass.sec}
The observing strategy and reduction steps of \hipass are described in
detail in Barnes et al. (2001). A full description of the galaxy
finding procedure and the source parametrization is given in Paper
I. Here we briefly summarize the \hipass specifics.

The observations were conducted in the period from 1997 to 2000 with
the Parkes\footnote{The Parkes telescope is part of the Australia
Telescope, which is funded by the Commonwealth of Australia for
operation as a National Facility managed by CSIRO.} 64-m radio
telescope, using the 21-cm multibeam receiver (Staveley-Smith et al.
1996).  The telescope scanned strips of $8^\circ$ in declination and
data were recorded for thirteen independent beams, each with two
polarizations.  A total of 1024 channels over a total bandwidth of 64
MHz were recorded, resulting in a mean channel separation of $\Delta
v=13.2~\kms$ and a velocity resolution of $\delta v=18~\kms$ after
Tukey smoothing. The data are additionally Hanning smoothed for
parameter fitting to improve signal-to-noise, giving a final velocity
resolution of $26.4~\kms$.  The total velocity coverage is $-1280$ to
$12700~\kms$.  After bandpass calibration, continuum subtraction and
gridding into $8^{\circ}\times 8^{\circ}$ cubes, the typical
root-mean-square (rms) noise is $13~\rm mJy\, beam^{-1}$.  This leads
to a $3\sigma$ column density limit of $\approx 6\times 10^{17} \rm
cm^{-2}$ per channel for gas filling the beam.  The spatial resolution
of the gridded data is $15\farcm5$.

The basic absolute calibration method used for \hipass is described by
Barnes et al. (2001).  The absolute flux scale was determined during
the first \hipass observations in February 1997 by calibrating a noise
diode against the radio sources Hydra A and 1934-638 with known
amplitudes (relative to the Baars et al. 1977 flux scale). The
calibration was checked regularly (on average three time each year) by
re-observing the two calibration sources. The r.m.s. of the flux
measurements averaged over all 13 beams and two polarizations is 2\%,
which gives a good indication of the stability of the absolute flux
calibration.

Two automatic galaxy finding algorithms were applied to the \hipass
data set to identify candidate sources.  To avoid confusion with the
Milky Way Galaxy and high velocity clouds, the range $v_{\rm
GSR}<300~\kms$ was excluded from the list.  The resulting list of
potential detections was subjected to a series of independent manual
checks.  First, to quickly separate radio frequency interference and
bandpass ripples from real \hi\ sources, two manual checks were done
examining the full detection spectra. Detections that were not
rejected by both checks were then examined in spectral, position,
RA-velocity, and dec-velocity space. Finally, the detections were
parametrized interactively using standard {\sc miriad} (Sault,
Teuben,\& Wright 1995) routines. This final catalogue of \hi-selected
sources is referred to as \hicat.

\section{Completeness} \label{completeness.sec}
The completeness $C$ of a sample is defined as the fraction of sources
from the underlying distribution that is detected by the survey.  For
an \hi-selected galaxy sample, $C$ is dependent on the peak flux,
\speak, and the velocity width, $W$, or alternatively on a combination
of both.

One way of determining the completeness is through analytical
methods. For example, for the AHISS sample presented in Zwaan et
al. (1997), a `detectability' parameter was calculated, which
depended on the distance of the detection from the center of the beam,
the variation of feed gain with frequency, the velocity width, and the
integrated flux. The completeness was assumed to be 100 per cent if the
detectability $>1$, corresponding to the requirement that \speak be
larger that 5 times the local rms noise level after optimal smoothing.
This analytically derived detectability was then compared with the
survey data and proved to be a satisfactory description of the survey
completeness.  Rosenberg \& Schneider (2002) used an empirical
approach to assess the completeness of their Arecibo Dual-Beam Survey
(ADBS), by inserting a large number of synthetic sources throughout
the survey data.  By determining the rate at which the synthetic
sources could be recovered, they established the completeness, which
they expressed as a function of signal-to-noise.

In this paper we also choose to assess the completeness of the \hipass
sample by inserting in the data a large number of synthetic sources,
prior to running the automatic galaxy finding algorithms (see Paper
I).  The actual process of source selection is a multi-step process,
which is partly automated and partly based on by-eye
verification. It is therefore preferable to study the completeness
empirically instead of analytically.

The synthetic sources were constructed to resemble real sources, and
were divided into three groups based on their spectral shapes:
Gaussian, double-horned, and flat-topped. The sources were not
spatially extended.  The velocity width, peak flux, and position of
each synthetic source were chosen randomly, and were drawn from a
uniform parent distribution that spans the range $20$ to $650~\kms$ in
$W$, the range $20$ to $130$ mJy in \speak, and the range $300$ to  
$10000~\kms$ in velocity.
  Care was taken not to place
synthetic sources on top of real sources. This was done by using the
results of an automatic galaxy finding algorithm that was run prior to
the insertion of the synthetic sources.  A total of 1200 synthetic
sources were inserted in the \hipass data cubes, with approximately
equal numbers of each of the three profile types.

\begin{figure*}
\center{\epsfxsize=0.8\hsize \epsfbox[0 270 550 568]{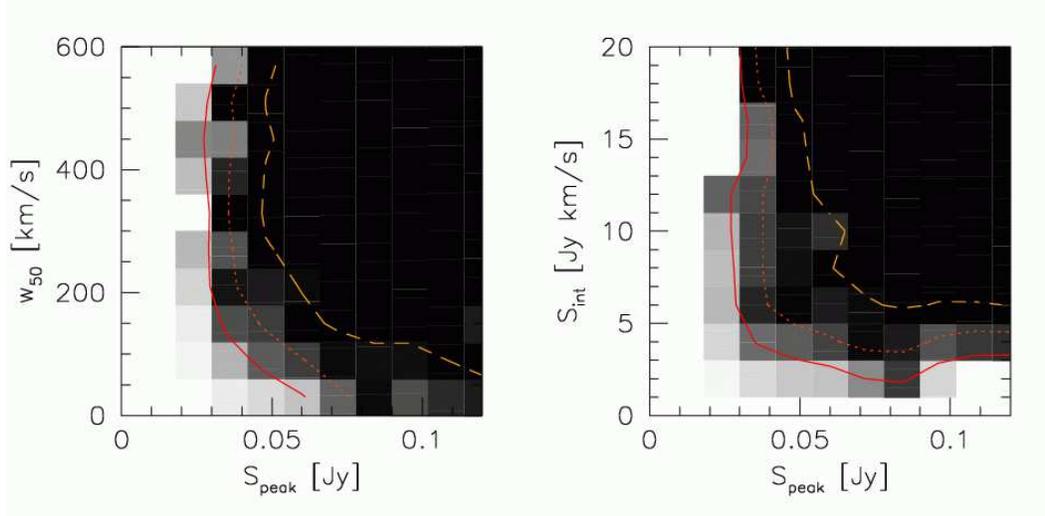}}
\caption{\label{2dcompl.fig} Bivariate completeness in the $\speakf,W$ plane
  and the $\speakf,\sint$ plane. Darker colors correspond to higher
  completeness. The contours indicate completeness levels of 50,
  75, and 95 per cent (from left to right).}
\end{figure*}

In Fig.~\ref{2dcompl.fig} we show a greyscale representation of the
completeness of the \hipass sample in the $\speakf,W$ plane and in the
$\speakf,\sint$ plane, where \speak\ is the peak flux density in Jy,
$W$ is velocity width in \kms, and \sint\ is the integrated flux in
\jykms.  The completeness in these plots is simply determined by
calculating $D$, the fraction of fake sources that is recovered in
each bin:
 \begin{equation}
 D(\speakf,W)=N^{\rm fake}_{\rm rec}(\speakf,W)/N^{\rm fake}(\speakf,W).
 \end{equation}

In order to calculate the completeness as a function of one parameter,
we need to integrate along one of the axes, and apply a weighting to
account for the varying number of sources in each bin.
Put differently, the completeness $C$ is the number of {\em detected}
real sources $N$ divided by the total number of {\em true} sources in each
bin, which we estimate with $N/D$. For example, the completeness as a
function of \speak, determined from the $\speakf,W$ matrix is given by
 \begin{equation}
 \label{compl.eq}
 C(\speakf)=\frac{\sum_{W=0}^{\infty}N(\speakf,W) }
 { \sum_{W=0}^{\infty}N(\speakf,W)/D(\speakf,W)}.
 \end{equation}
This weighting corrects for the fact that the parameter distribution
of the synthetic sources might be different from that of the
underlying real galaxy distribution.  Similarly, $C(W)$ can be
determined by integrating over \speak, and $C(\sint)$ can be determined
by integrating over a $\speakf,\sint$ matrix. Hereafter, we refer to
$C$ as the differential completeness since it refers to the
completeness at a certain value of \speak, \sint, or $W$.

It is often convenient to calculate the cumulative completeness
$C^{\rm cum}$. For example, $C^{\rm cum}(\speakf)$ is the completeness for
all sources with peak fluxes larger than \speak: 
 \begin{equation}
 C^{\rm cum}(\speakf)=\frac{\sum_{\speakf'=\speakf}^{\speakf'=\infty}\sum_{W=0}^{\infty}N(\speakf',W) }
 {\sum_{\speakf'=\speakf}^{\speakf'=\infty}\sum_{W=0}^{\infty}N(\speakf',W)/D(\speakf',W)}.
 \end{equation}

\begin{figure*}
\center{\epsfxsize=17.5cm \epsfbox[18 470 592 718]{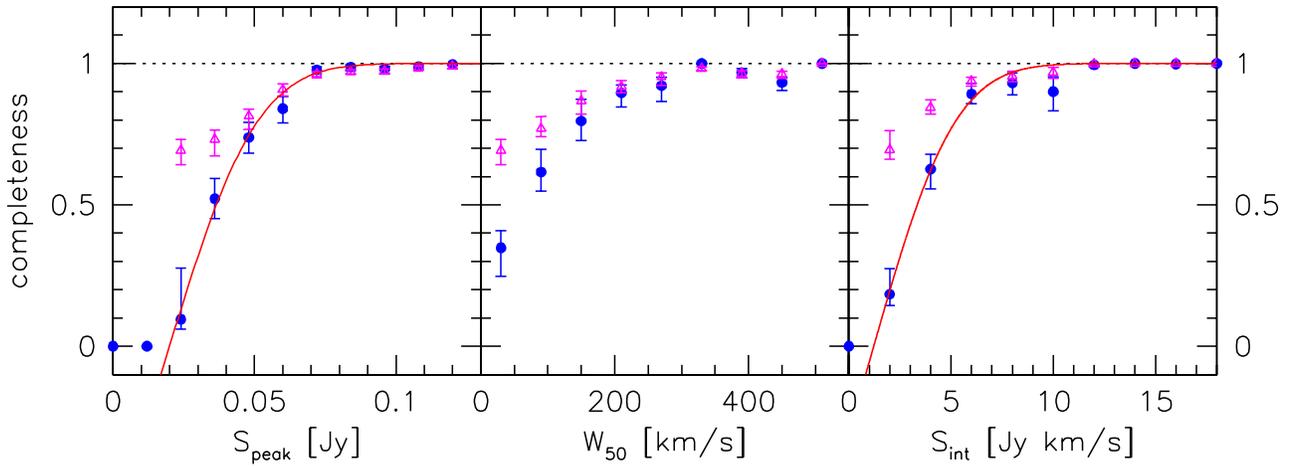}}
\caption{\label{completeness.fig} Completeness of \hicat as measured
  from the detection rate of synthetic sources. Solid circles show the
  differential completeness and triangles the cumulative
  completeness. The solid lines are error function fits to the points,
  with the fit parameters given in Table~\ref{completeness.tab}.
  Error bars indicate 68 per cent confidence levels.}
\end{figure*}

\subsection{Results} \label{complresuls.sec}
Fig.~\ref{completeness.fig} shows the result of this analysis, the
circles show the differential completeness as a function of \speak,
$W_{50}$, and \sint, and the triangles show the cumulative
completeness. Error bars indicate 68 per cent confidence levels and
are determined by bootstrap re-sampling\footnote{From the parent
population of $N$ synthetic sources, $N$ sources are chosen randomly,
with replacement. This is repeated 200 times and for each of these 200
re-generated samples the completeness $C'$ is calculated following
equations 2 and 3.  The $1\sigma$ upper and lower errors on the
completeness are determined by measuring from the distribution of $C'$
the 83.5\% and 16.5\% percentiles}.  We fit the completeness as a
function of \speak and \sint\ with error functions (erf), which are
indicated by solid lines.  The best-fitting error functions are given
in Table~\ref{completeness.tab}, along with the completeness limits at
95 and 99 per cent.

\begin{center}
 \begin{table*}
 \caption[]{Completeness}
 \label{completeness.tab}
 \begin{tabular}{llccl}
 \hline
{parameter} & {completeness} & {$C=0.95$} & {$C=0.99$} \\
\hline
\speak (mJy)       & ${\rm erf}[0.028(\speakf-19)]$       &  68    & 84  \\
$\sint$ (\jykms)& ${\rm erf}[0.22(\sint-1.1)]$   &  7.4  & 9.4 \\
\hline
$\speakf({\rm mJy}),\sint$(\jykms) & \multicolumn{3}{l}{${\rm erf}[0.036(\speakf-19)] {\rm erf}[0.36(\sint-1.1)]$}  \\
\noalign{\smallskip} 
\hline
 \end{tabular}
 \medskip
\end{table*}
\end{center}

Clearly, there is not a sharp segregation between detectable and not
detectable for any of the three parameters under examination. The
completeness is a slowly varying function, which illustrates the
complexity of the detectability of \hi\ signals.  However, all curves
reach the 100 per cent completeness level. This indicates that our source
finding algorithms do not miss any high signal-to-noise sources, and
our system of checking all potential sources for possible confusion
with RFI is sufficiently conservative that it does not cause many
false negatives.

\begin{figure*}
\center{\epsfxsize=0.8\hsize \epsfbox[18 420 592 718]{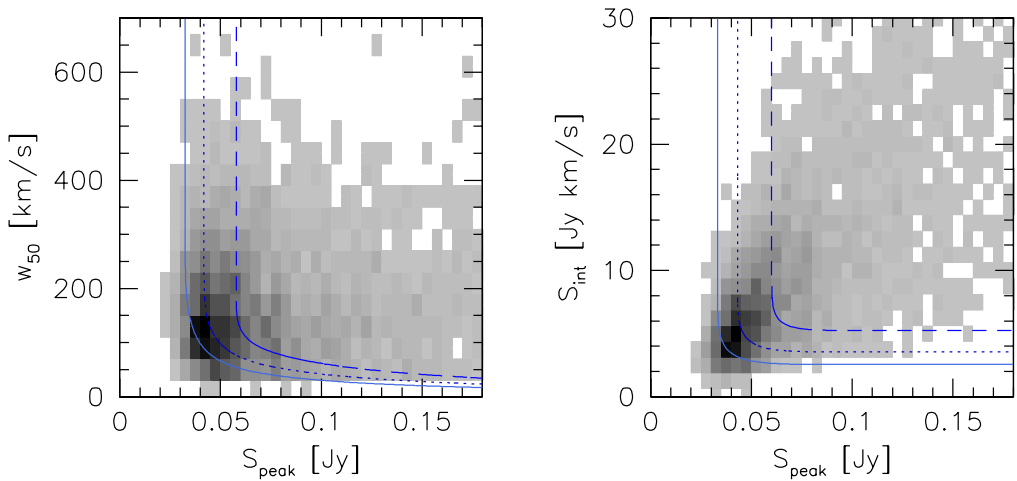}}
\caption{\label{2ddist.fig} Bivariate parameter distribution of
  \hicat. Darker colors correspond to higher source densities.
  Analytical approximations of the completeness limits at 50, 75,
  and 95 per cent (from left to right) are indicated by curves.  }
\end{figure*}

Although the above derived expressions are useful for understanding
the completeness of \hicat, they do not allow us to calculate
completeness levels for individual sources. For many purposes, for 
example in evaluating the \hi\ mass function, it is convenient to know
what the completeness of the catalogue is for a source with specific
parameters. We tested different fitting functions and found that the
completeness can be fitted satisfactorily using two parameters:
\begin{equation}
\label{complexpr.eq}
C(\speakf,\sint)={\rm erf}[0.036(\speakf-19)] {\rm erf}[0.36(\sint-1.1)].
\end{equation} 
 This provides an accurate fit to the completeness matrices shown in
Fig.~\ref{2dcompl.fig}, and also reproduces the one-parameter fits
shown in Fig.~\ref{completeness.fig}, after the proper summation given
Eq.~\ref{compl.eq} has been applied. In Fig.~\ref{2ddist.fig} the 50,
75, and 95 per cent completeness limits calculated using
Eq.~\ref{complexpr.eq} are drawn on top of the parameter distribution
of the full \hicat data. The contours in the $\speakf,W$-plane are
calculated by assuming $W=1.22 \sint/\speakf$, which provides a good
fit to the data.
Unfortunately, the regions of parameter space
that are most densely populated are severely incomplete, as is
generally true for samples that do not have a sharp completeness
limit.  By cutting \hicat at the 95 (99) per cent completeness limit,
the sample is reduced to 2209 (1678) sources.

\subsection{Verification of completeness limits}
Among blind extragalactic \hi\ surveys, \hipass is unique in the sense
that it is fully noise-limited. Surveys such as AHISS or the ADBS are
partly bandwidth-limited, which means that the brightest galaxies in
the sample can only be detected out to the distance limit set by the
restricted bandwidth of the receiving system. Since \hipass is a
relatively shallow survey and was conducted with a large bandwidth (64
MHz), even the detection of the most \hi\ massive galaxies is
noise-limited.  The distance distribution $N(D)$ of \hicat galaxies
drops to zero at large distances, before the maximum distance of
$12,700~\kms$ is reached (see paper I).  This property of \hicat
enables the use of standard techniques to verify the completeness
limits determined in Section~\ref{complresuls.sec}.  For
bandwidth-limited samples these methods would not give meaningful
results.

\begin{figure}
\center{\epsfxsize=8cm \epsfbox[18 150 592 718]{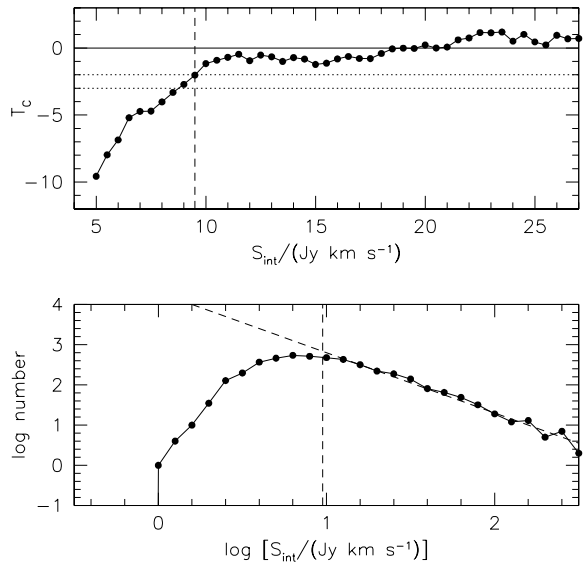}}
\caption{\label{rauzy.fig} Test of completeness limits in \hicat. The
  top panel shows the $T_C$ estimator (see text) as a function of
  integrated flux. The completeness limit is reached at
  $\sint=9.5~\jykms$, where $T_C=-2$. The bottom panel shows the
  number of sources as a function of \sint. The fitted line
  corresponds to the expected distribution $dN\propto S_{\rm
  int}^{-5/2}d\sint$ for a flux-limited sample.  The completeness limit of
  $\sint=9.5~\jykms$ is indicated by a vertical dashed line.}
\end{figure}

Rauzy (2001) recently suggested a new tool to assess the completeness
for a given apparent magnitude in a magnitude-redshift sample. This
method is easily adapted to an \hi-selected galaxy sample.
Essentially, the method compares the number of galaxies brighter and
fainter than every galaxy in the sample. In the case of a
homogeneously distributed sample in space, the method is essentially
the same as a $V/V_{\rm max}$ test, but by design Rauzy's method is
insensitive to structure in redshift space. The method is based on the
definition of a random variable $\zeta$, which for a \hi-selected
sample can be defined as
\begin{equation}
\zeta=\frac{\Theta(M_{\rm HI})}{\Theta[M^{\rm lim}_{\rm HI}(Z)]},
\end{equation}
where $\Theta$ is the cumulative \hi\ mass function, $Z$ is a
`distance modulus' defined as $Z=\log \sint-\log\mhi$, and $M^{\rm
lim}_{\rm HI}(Z)$ is the limiting \hi\ mass at the distance
corresponding with $Z$.  An unbiased estimate of $\zeta$ for object
$i$ is given by
\begin{equation}
\zeta_i=\frac{r_i}{n_i+1},
\end{equation}
 where $r_i$ is the number of objects with $\mhi\ge \mhi_i$ and
$Z\le Z_i$, and $n_i$ is the number of objects for which $\mhi\ge \mhi^{\rm
lim}_i(Z_i)$ and $Z\le Z_i$.  The values of $\zeta_i$ should be uniformly distributed
between 0 and 1.
Now a quantity $T_C$ can be defined as 
\begin{equation}
T_C=\frac{ \sum_{i=1}^{N_{\rm gal}}(\zeta_i-1/2)}{(\sum_{i=1}^{N_{\rm gal}}V_i)^{1/2}},
\end{equation}
 where $V_i$ is the variance of $\zeta_i$, defined as
$V_i=(n_i-1)/[12(n_i+1)]$.  The completeness of the sample can now be
estimated by computing $T_C$ on truncated subsamples according to a
decreasing \sint. For statistically complete subsamples the quantity
$T_C$ has an expectation value of zero and unity variance. The
completeness limit is found when $T_C$ drops systematically to
negative values, where $T_C=-2\, (-3)$ indicates a 97.7 (99.4) per cent
confidence level.  In the top panel of Fig.~\ref{rauzy.fig} we plot
the result of the $T_C$ completeness test. From this we derive that
the completeness limit of the sample is $S_{\rm int}^{\rm
lim}=9.5~\jykms$ at the 97.7 per cent confidence level.  
This limit
is very close to what was found in the previous section, where we
calculated the completeness based on the detection rate of synthetic
sources.

As a final verification we plot in the bottom panel of
Fig.~\ref{rauzy.fig} the number of galaxies as a function of
\sint. The dotted line shows a $dN\propto S_{\rm int}^{-5/2}d\sint$
distribution expected for a flux-limited sample, and is scaled
vertically so as to fit the right hand side of the curve.  Deviations
from the curve start to become apparent at $\sint=10~\jykms$, which is
consistent with the more accurate determination from the $T_C$
method. Unlike the $T_C$ method, this method of plotting $dN$ as a
function of \sint\ is sensitive to the effects of large scale
structure.

\subsection{Completeness as a function of sky position}
\begin{figure}
\center{\epsfxsize=8.0cm \epsfbox[70 160 550 700]{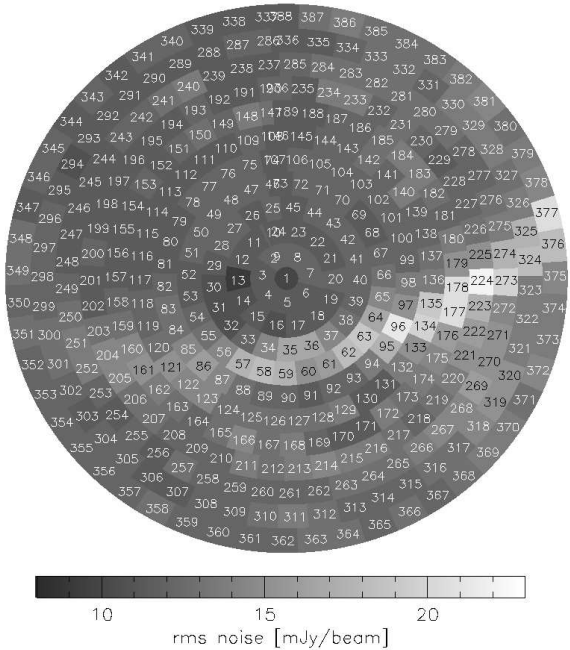}}
\caption{\label{noise_on_sky.fig} Median noise levels in the southern
  \hipass cubes. The south celestial pole is in the center, RA=0 is on
  top, and increases counter-clockwise.  The scale bar shows the noise
  levels in $\rm mJy\,beam^{-1}$.  The horizontal bright band corresponds
  to $b=0$, where the noise level is elevated. The numbers correspond to 
  the numbers of the $8^\circ\times8^\circ$ \hipass cubes.}
\end{figure}

\hipass achieves 100 per cent coverage over the whole southern sky and
has a mostly uniform noise level of $13.0\,\rm
mJy\,beam^{-1}$. However, in some regions of the sky the noise level
is elevated due to the presence of strong radio continuum sources. In
Fig.~\ref{noise_on_sky.fig} the median noise level of every \hipass
cube is shown. These noise levels are determined robustly using the
estimator $\sigma=s(\pi/2)^{1/2}$, where $s$ is the median absolute
deviation from the median. This estimator is much less sensitive to
outliers than the straight rms calculation, and provides an accurate
estimate of the rms of the underlying distribution, provided that this
distribution is nearly-normal.  The average cube noise level is
elevated more than 10 per cent over just 14.8 per cent of the sky, and
elevated more than 20 per cent over 6.2 per cent of the sky.  A region
of elevated noise levels can be clearly identified in
Fig.~\ref{noise_on_sky.fig}, where the highest noise values go up to
$22\,\rm mJy\,beam^{-1}$. This region corresponds very closely to
Galactic Plane, where the strongest radio continuum sources are
located and where the density of continuum sources is highest.

It is not straightforward to assess accurately how the
completeness is affected by varying noise levels. Since a
significantly different noise level is only observed over a small
region of the sky, the number of synthetic sources in these regions is
too small to calculate the completeness limits
accurately. Furthermore, the region of highest noise levels
coincidently lies in the direction of the Local Void, where the
detection rate of sources is naturally depressed. Therefore, the $T_C$
method, or a simple $V/V_{\rm max}$ method are also unreliable
estimators of the completeness here.  In the absence of empirical
estimators, we make the reasonable assumption that the detection
efficiency scales linearly with the local noise level, which means
that the completeness $C(\speakf)$ can be replaced with $C(\speakf\times
13.0/\sigma)$ in regions of atypical noise levels. This implies that
the 95 per cent completeness level, which is normally reached at 71~mJy, is
reached at 85~mJy when the noise level is elevated by 20 per cent. The
completeness as a function of $W_{50}$ is probably not affected by a
slight increase in noise level. The completeness as a function of
\sint\ is adjusted similar to $C(\speakf)$.

\subsection{Completeness as a function of profile shape}
\begin{figure*}
\center{\epsfxsize=17.5cm \epsfbox[18 470 592 718]{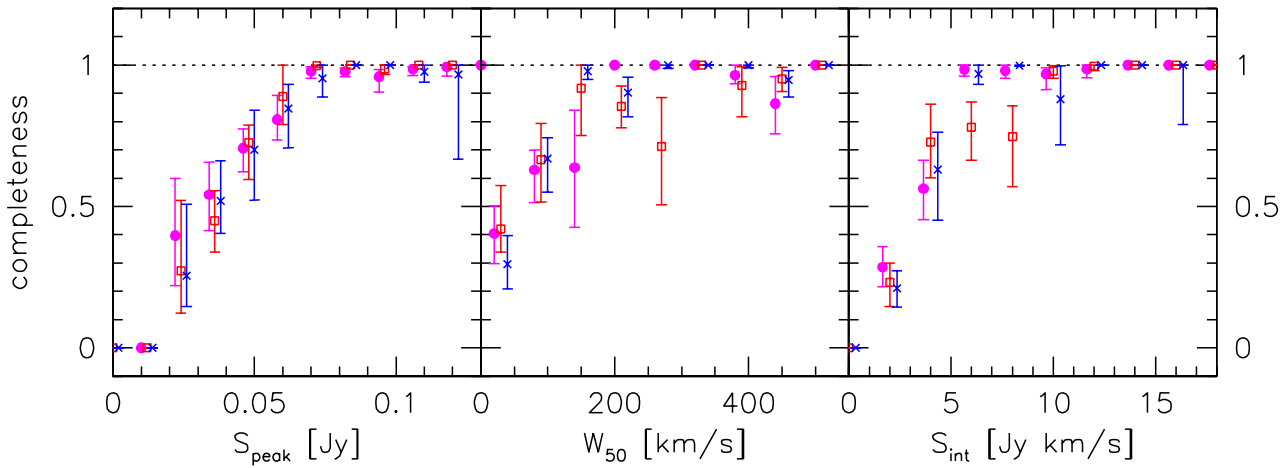}}
\caption{\label{comp_prof.fig} Completeness of \hicat as a function of
  profile shape.  Solid circles indicate Gaussian profiles, open boxes
  double-horned profiles, and crosses flat-topped profiles. The points
  for the Gaussian and flat-topped profiles are offset horizontally to
  avoid overlapping of points.  }
\end{figure*}

In order to test the detection efficiency of various profile shapes,
the synthetic sources were divided into three groups: Gaussian,
double-horned, and flat-topped. We perform the completeness analysis
for each of these subsamples individually, and show the results in
Fig.~\ref{comp_prof.fig}.  Within the errors, the detection
efficiency as a function of peak flux is independent of profile shape.
However, $C(W_{50})$ and $C(\sint)$ are
somewhat depressed for double-horned profiles with respect to Gaussian
and flat-topped profiles. The reason for this is probably that low
signal-to-noise double-horned profiles are easily mistaken for two
noise peaks, whereas Gaussian and flat-topped profiles have their flux
distributed over adjoining channels, which together stand out from the
noise more clearly.

\subsection{Completeness as a function of velocity}


In Fig.~6 of Paper I we show that the velocity distribution of the
initial sample of potential \hicat detections shows strong peaks at
known RFI frequencies and frequencies corresponding to hydrogen
recombination lines. This might give rise to the concern that the
completeness of \hicat is suppressed at these frequencies. However, in
Paper I we show that the three-dimensional signature of these
contaminating signals is sufficiently characteristic that they can be
reliably removed from the catalogue. The final distribution of \hicat
velocities shows no features that correlate with RFI or hydrogen radio
recombination line frequencies, indicating that the completeness is
not significantly affected at these frequencies. Unfortunately, we are
not able to further substantiate this claim since 1200 uniformly
distributed synthetic sources provide insufficient velocity sampling
to study the completeness as function of velocity in detail.


\section{Reliability}
\label{reliability.sec}
 The reliability of the sample was determined by re-observing a
subsample of sources with the Parkes Telescope. The aim of the
observations was twofold: assessing the reliability of \hicat as a
function of peak flux, integrated flux and velocity width, and
removing spurious detections from the catalogue. The subsample was
chosen in such a way that the full range of \hicat parameters is
represented, but preference was given to those detections that have
low integrated fluxes. For every observing session, a sample was
created that consisted of randomly chosen \hicat detections,
complemented with detections with low \sint (generally lower than
8~\jykms). At the time of the observations, the observer chose
randomly from these samples.  The full range of RA was covered by the
observations.

The observations were carried out over five observing sessions between
September 2001 and November 2002. They were done in narrow-band mode,
which gives 1024 channels over 8 MHz, resulting in a spectral
resolution of $1.65~\kms$ at $z=0$. In this narrow-band correlator
setting only the inner 7 beams of the multibeam system are available.
An observing mode was used where the target is placed sequentially in
each of the 7 beams and a composite off-source spectrum is calculated
from the other 6 beams.  This strategy yields a noise level $1.85$
times lower than standard on-off observations in the same amount of
time.  Typical integration times were 15 min.  The narrow-band
observations yield lower rms noise levels than standard broad-band
multibeam observations.  Furthermore, the high frequency resolution
enables better checks of the reality of \hicat sources since narrow
signals can be detected in several independent channels.  The data
were reduced using the {\sf AIPS++} packages {\sc livedata} and {\sc
gridzilla} (Barnes et al. 2001), and the detections were parametrized
using standard {\sc miriad} routines.

\begin{figure*}
\center{\epsfxsize=17.5cm \epsfbox[18 470 592 718]{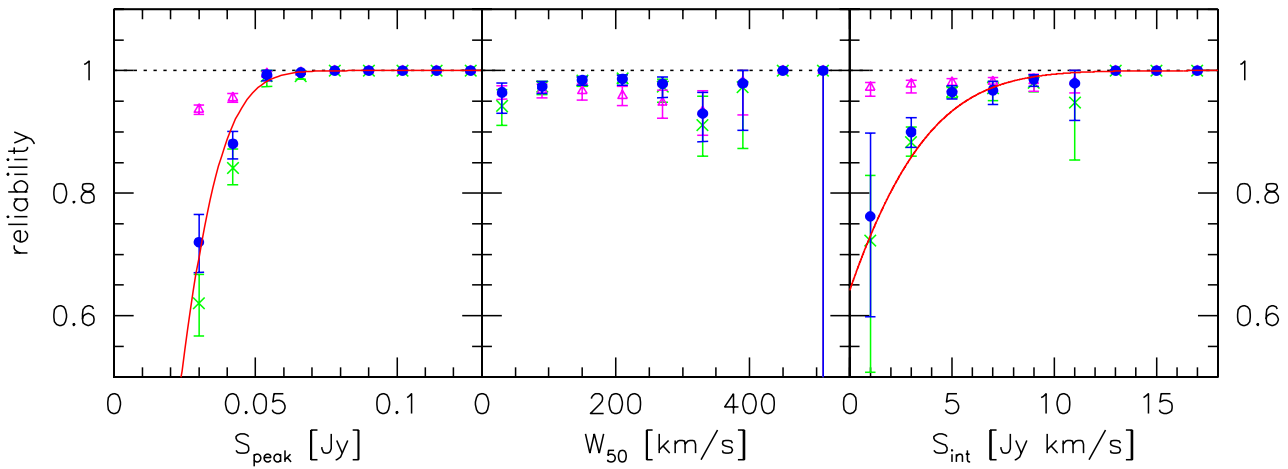}}
\caption{\label{reliability.fig} Reliability of \hicat measured from
  Parkes follow-up observations. Solid circles show the differential
  reliability, triangles the cumulative reliability. The crosses show
  the reliability of \hicat before unconfirmed sources have been taken
  out. The solid lines are error function fits to the points, with the
  fit parameters given in Table~\ref{reliability.tab}.  Error bars
  indicate 68 per cent confidence levels.}
\end{figure*}

First, we consider the reliability of the original catalogue, before
unconfirmed sources have been taken out. 
The fraction of sources that was confirmed is defined as
\begin{equation}
T(\speakf,W)=N^{\rm rel}_{\rm conf}(\speakf,W)/N^{\rm rel}_{\rm obs}(\speakf,W), 
\end{equation}
where $N^{\rm rel}_{\rm conf}$ and $N^{\rm rel}_{\rm obs}$ are the
number of confirmed and observed sources, respectively.  The
reliability as a function of peak flux \speak is the mean
of $T$, weighted by the number of sources in each bin:
 \begin{equation} \label{rel.eq}
 R(\speakf)=\frac{\sum_{W=0}^{\infty}N(\speakf,W)\times T(\speakf,W)}
 { \sum_{W=0}^{\infty}N(\speakf,W)},
 \end{equation}
and the cumulative reliability is
 \begin{equation} \label{relcumu.eq}
 R^{\rm cum}(\speakf)=
 \frac{\sum_{\speakf'=\speakf}^{\speakf'=\infty}\sum_{W=0}^{\infty}N(\speakf',W)\times T(\speakf',W)}
 {\sum_{\speakf'=\speakf}^{\speakf'=\infty} \sum_{W=0}^{\infty}N(\speakf',W)}.
 \end{equation}
Again, analogous methods can be used to measure $R(W)$ and $R(\sint)$.
Fig.~\ref{reliability.fig} shows the measured reliability 
as a function of \speak, $W_{50}$, and \sint. The crosses show the
differential reliability, error bars indicate 68 per cent confidence levels
and are determined by bootstrap re-sampling the data 200 times.

As sources that were re-observed but not confirmed were taken out of
the catalogue, by re-observing a subsample of sources we improve the
catalogue reliability. Eventually if we were to re-observe all sources
the reliability would rise to 100 per cent. To calculate the reliability
after taking out unconfirmed sources, we have to estimate the {\em
expected\/} number of real sources, which is the number of confirmed
sources plus $T$ times the number of sources that have not been
observed.

A final complication arises because a second subsample of sources from
\hicat was re-observed as part of a program to measure accurate velocity
widths (Meyer et al. 2004b). This program also influences the
reliability because non-detections were taken out of the catalogue and
detections are marked as `confirmed' in \hicat. This latter class of
sources is indicated as $N^{\rm NB}_{\rm conf}$. Now, 
the expected number of real sources is given by
 \begin{equation}
 N_{\rm exp real}=N^{\rm rel}_{\rm conf} + N^{\rm NB}_{\rm conf}+(N - N^{\rm rel}_{\rm conf} - N^{\rm NB}_{\rm conf}) \times (N^{\rm rel}_{\rm conf}/N^{\rm rel}_{\rm obs}).
 \end{equation}
Note that the total number of sources in \hicat, $N$, excludes all
unconfirmed sources. Now, we can redefine $T$ as
\begin{equation}
T(\speakf,W)=N_{\rm exp real}(\speakf,W)/N(\speakf,W), 
\end{equation}
and equations~\ref{rel.eq} and \ref{relcumu.eq} can be used to
calculate the reliability of \hicat. The circles and triangles in 
Fig.~\ref{reliability.fig} show the measured differential and
cumulative reliability, respectively. In total, 1201 sources were
observed, of which 119 were rejected.

\subsection{Results} 
The overall reliability is very high (95 per cent), partly because the
catalogue was cleaned up considerably by re-observing many sources and
rejecting unconfirmed sources from the catalogue. The reliability
drops significantly below $\speakf<50$~mJy and $\sint <5\,\rm Jy
\,km\,s^{-1}$, and there is possibly a reduced reliability around
$W_{50}=350~\kms$. This latter feature may be related to the confusion
of real \hi\ emission signals with ripples in the spectral passband.
We fit the reliability as a function of peak flux and integrated flux
with error functions, the parameters of which are presented in
Table~\ref{reliability.tab}.  The 99 per cent reliability level is reached at
$\speakf=58~\rm mJy$ and $\sint=8.2~\jykms$.  If sources with a \hicat
comment `2=have concerns' are removed from the sample, the overall
reliability rises to 97 per cent. Similarly to the results found for the
completeness levels, we find that the reliability of individual
sources can be determined satisfactorily as a function of \speak and
\sint. The functional form is given in Table~\ref{reliability.tab}.

\begin{center}
 \begin{table*}
 \caption[]{Reliability}
 \label{reliability.tab}
 \begin{tabular}{llccl}
 \hline
{parameter} & {reliability} & {$C=0.95$} & {$C=0.99$} \\
\hline
\speak (mJy)       & ${\rm erf}[0.040(\speakf-12)]$       &  50    & 58  \\
$\sint$ (\jykms)& ${\rm erf}[0.12(\sint+6.4)]$   &  5.0  & 8.2 \\
\hline
$\speakf({\rm mJy}),\sint$(\jykms) & \multicolumn{3}{l}{${\rm erf}[0.045(\speakf-12)] {\rm erf}[0.20(\sint+6.4)]$}  \\
\noalign{\smallskip} 
\hline
 \end{tabular}
 \medskip
\end{table*}
\end{center}

\section{Parameter uncertainties} \label{parameter.sec}
A detailed description of all measured parameters in \hicat is
presented in Paper I. Here we discuss the error estimates of the most
important parameters: peak flux, integrated flux, velocity width,
heliocentric recessional velocity ($cz$) and sky position.  Other
authors have discussed analytical approaches to estimating
uncertainties on \hi\ 21-cm parameters (e.g., Schneider et al. 1990,
Fouqu{\'e} et al. 1990, Verheijen \& Sancisi 2001), but for \hicat\
sufficient comparison data are available to measure the errors
empirically.  In this analysis we make use of the synthetic source
parameters and the narrow-band observations to determine the total
observational errors on the parameters.
The data published in the
\hipass BGC (Koribalski et al. 2004) are used to establish what
fraction of the error is caused by the parametrization procedure.

We assume that the error $\sigma_X$ on
parameter $X$ can be satisfactorily described by 
\begin{equation}
\label{error.eq}
\sigma(X)= c_1 Y^{n}+c_2,
\end{equation}
where $Y$ is a parameter that can be equal to $X$ or any other
parameter, and $n$, 
$c_1$, and $c_2$ are constants. There is no physical basis for this
analytical description of the errors, but we find later that 
Eq.~\ref{error.eq} provides satisfactory fits to the measured
parameter uncertainties. In the following we determine how each 
$\sigma(X)$ depends on all parameters.

When comparing parameters from different data sets, we know that the
measured rms scatter on the difference between \hicat parameter $X$ and
parameter $X$ from data set $Z$ is given by
\begin{equation}
\label{sigma1.eq}
\sigma(X)_{\rm meas}^2=\sigma(X)_{\rm HICAT}^2+\sigma(X)_{\rm Z}^2,
\end{equation}
where $\sigma(X)_{\rm meas}$ is the measured rms scatter 
on $X_{\rm HICAT}-X_{\rm Z}$, $\sigma(X)_{\rm HICAT}$ is the
error in the \hicat parameter, and $\sigma(X)_{\rm Z}$ is the
error in data set $Z$. The latter two parameters are unknown,
but we can make the simplifying assumption that 
\begin{equation}
\label{sigma2.eq}
\sigma(X)_{\rm Z}=\sigma(X)_{\rm HICAT} \frac{\rm rms_{\rm Z}}{\rm rms_{\rm HICAT}}
\end{equation}
where $\rm rms_{\rm Z}$ denotes the rms noise in the survey on which 
catalogue $Z$ is based.

\begin{figure}
\center{\epsfxsize=10cm \epsfbox[132 167 618 714]{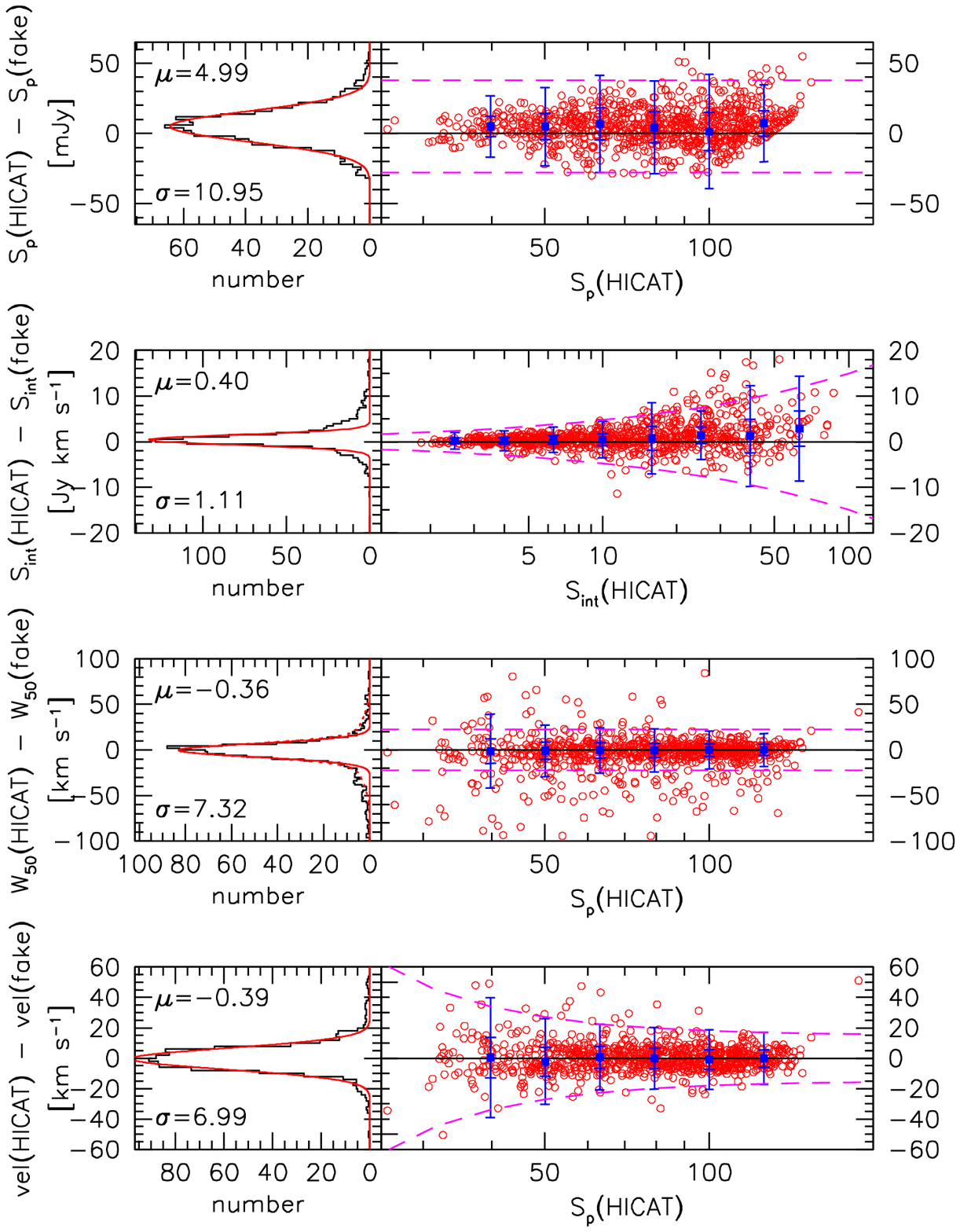}}
\caption{\label{fakeparams.fig} Comparison between \hicat and synthetic
source parameters. The left panels show the histograms of the
differences, which are fitted by Gaussian profiles for which the
parameters are indicated in the top left corners. The right panels show
the differences as a function of the measured \hicat parameters. The
points and error bars show the zero-points and the widths of the fitted
Gaussians (inner and outer error bars indicate $1\sigma$ and $3\sigma$,
respectively). The dashed lines are the best-fitting analytical
descriptions of $3\sigma(X)$.}
\end{figure}

\subsection{Error estimates from comparison with synthetic sources}
First, we compare the \hicat parameters with those of the synthetic
sources. This comparison is particularly useful for estimating errors
because the synthetic source parameters are noise free, which means
that $\sigma(X)_{\rm fake}=0$ for all parameters. Therefore, the
measured $\sigma(X)_{\rm meas}$ is equal to $\sigma(X)_{\rm HICAT}$,
which is the parameter of interest.  In Fig.~\ref{fakeparams.fig} we
plot the difference between the measured \hicat parameters and
parameters of the synthetic sources that were inserted into the
data. The left panels show the difference histograms, fitted by
Gaussian profiles. Parameters for these are indicated in the top left
corners. The right panels show the differences as a function of the
measured \hicat parameters. The points and error bars show the
zero-points and widths of the fitted Gaussians (inner and outer
error bars indicate $1\sigma$ and $3\sigma$, respectively) in different
bins. We prefer Gaussian fitting to calculating straight rms values
because this latter estimator is much more sensitive to outliers.  In
the right panels we indicate the best-fitting relations for $3\sigma(X)$
by dashed lines.

As is expected, the error on \speak is independent of peak flux. The
measurement error is just determined by the 13.0~mJy rms noise in the
spectra, but lowered to a measured value of 11.0~mJy due to the fact
that more than one channel may contribute to the measurement of
\speak. The error on \speak is not found to be dependent on any of the other
parameters, so we adopt a fixed value of $\sigma(\speak)=11.0\, \rm mJy$.
The other effect than can be seen in this panel is that there is a
global offset of 5.0 mJy in the measured \speak with respect to the peak
flux of the synthetic sources.  This effect, which in Zwaan et
al. (2003) is referred to as the `selection bias', arises because
after adding noise to a spectrum the measured peak flux density is
generally an overestimation of the true peak flux density. 

The error in \sint\ is found to be dependent on \sint\ only, and can
be satisfactorily fitted with $c_1=0.5$, $n=1/2$.  This implies that
$\sigma(\sint)=1.5~\jykms$ (or 16 per cent) at the 99 per cent
completeness limit of $9.4~\jykms$. Fouqu{\'e} et al. (1990) derive
that $\sigma(\sint)$ is dependent on both \sint\ and \speak\ as
$\sigma(\sint)\propto \sint^{1/2} \speak^{-1/2}$. Our analysis shows
that for the \hicat\ data $\sigma(\sint)$ can be described
satisfactorily as a function of \sint\ only.  The error in \sint\ will
be the dominant factor in the error on the \hi\ mass, except for the
nearest galaxies for which peculiar velocities contribute
significantly to the uncertainty in \hi\ mass.

The error in $W_{50}$ is not clearly dependent on any other parameter,
so we adopt a constant $\sigma(W_{50})=7.5\, \rm km s^{-1}$.  It
should be noted, however, that there appears to be an excess of points
which is not satisfactorily fitted with a single Gaussian.  These outliers
are preferentially those with low peak fluxes, but large velocity
widths. Larger uncertainties in the measurements of 
velocity width occur with broad, low signal-to-noise profiles, because
the edges of the profiles can not always be chosen unambiguously.
Approximately one-third of the measurements can be fitted with a Gaussian 
with $\sigma=25\,\rm km\,s^{-1}$.

We find that the error on recessional velocity is dependent on \speak
only, with higher peak flux detections having lower errors on the
measured $V_{\rm hel}$. The error bars can be fitted with parameters
$c_1=1.0\times 10^4$, $n=-2$ and $c_2=5$. Fouqu{\'e} et al. (1990)
find for their data that $n=-1$, but incorporate an additional
dependence on the steepness of the \hi\ profile.

\begin{figure}
\center{\epsfxsize=10cm \epsfbox[132 292 618 714]{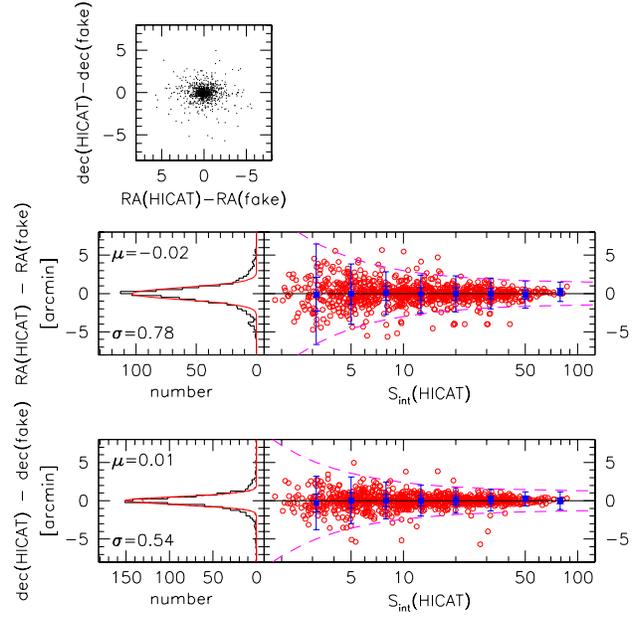}}
\caption{\label{posdiff.fig} The top panel shows the 
difference between the position of the inserted synthetic sources and the
fitted position after parametrization, in arcmin. 
The lower panels are similar to
those in Fig.~\ref{fakeparams.fig}.}
\end{figure}

In Fig.~\ref{posdiff.fig} the top panel shows the difference between
position of the inserted synthetic sources and the fitted position
after parametrization.  The lower two panels show the position
differences as a function of \sint. The positional accuracy in RA is
fitted with $c_1=5.5$, $n=-1$, $c_2=0.45$, and the accuracy in Dec is
fitted with $c_1=4$, $n=-1$, $c_2=0.4$.  These numbers imply that the
positional accuracy at the 99 per cent completeness limit is
$1\farcm05$ in RA, and $0\farcm82$ in Dec.  The difference between
these two numbers arises because the \hipass data are more regularly
sampled in the Dec direction (see Barnes et al. 2001).  This
positional accuracy agrees very well with the results found from
\hicat matching with the 2MASS Extended Source Catalogue (Jarret et
al. 2003, see Meyer et al. 2004b).

 \begin{table}
 \caption[]{Parameter uncertainties}
 \label{errors.tab}
 \begin{tabular}{lll}
 \hline
{parameter} & error & $\sigma$ at $C=0.99$  \\
\hline
$\sigma(\speakf)$           & $11.0\,\rm mJy$   & $11.0\,\rm mJy$\\
$\sigma(\sint)$       & $0.5 S_{\rm int}^{1/2}\,\jykms$ & $1.5 \,\jykms$ \\
$\sigma(W_{50})$      & $7.5\,\rm km\,s^{-1}$ & $7.5\,\rm km\,s^{-1}$\\
$\sigma(V_{\rm hel})$ & $1.0\times 10^4 \speakf^{-2}+5\,\rm km\,s^{-1}$ & $6.4\, \rm km\,s^{-1}$\\
$\sigma(\rm RA)$      & $5.5 S_{\rm int}^{-1}+0.45\, \rm arcmin$ & $1.05\, \rm arcmin$\\
$\sigma(\rm dec)$     & $4 S_{\rm int}^{-1}+0.4\, \rm arcmin$ & $0.82\, \rm arcmin$\\
\noalign{\smallskip} 
\hline
 \end{tabular}
 \medskip
\end{table}

\subsection{Verification of error calculations with Parkes follow-up observations}
Although the comparison with noise-free parameters in the previous
subsection is a useful method of estimating the errors on \hicat
parameters, it is important to verify these results with independent
measurements. Such measurement are available through our program of
Parkes narrow-band (NB) follow-up observations, which was described in
Section~\ref{reliability.sec}. These follow-up observations are
preferentially targeted at sources with low integrated fluxes, but the
sample is sufficiently large to make a meaningful parameter comparison
over a large dynamic range. The NB observations were carried out
independently from the \hipass program and consisted of pointed
observations instead of the active scanning used for \hipass.
The spectral resolution of the NB observations was 1.65~\kms, compared
to 13.2~\kms\ for \hipass, but the data used in this section were
smoothed to the \hipass resolution. The NB profiles were parametrized
with the same {\sc miriad} software used for \hicat.

\begin{figure}
\center{\epsfxsize=10cm \epsfbox[132 167 618 714]{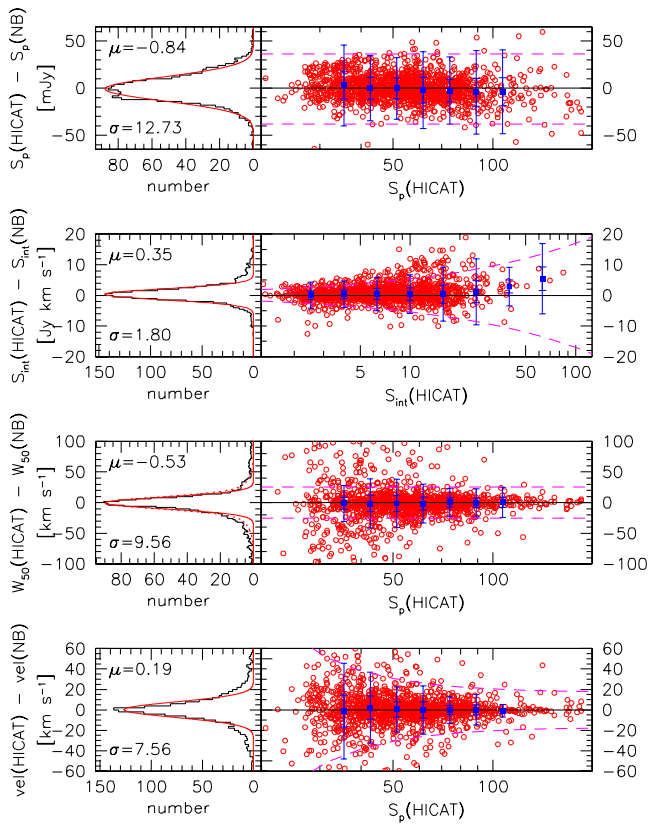}}
\caption{\label{NBparams.fig} Similar to Fig.~\ref{fakeparams.fig},
  but showing the comparison between parameters from \hicat and
  narrow-band follow-up observations.}
\end{figure}

In Fig.~\ref{NBparams.fig} the differences
between \hicat parameters and those from follow-up observations is
presented. The dashed lines in the right-hand panels are not fits to
the error bars, but are the equations given in Table~\ref{errors.tab},
converted using Eq.~\ref{sigma1.eq} and Eq.~\ref{sigma2.eq}. Here we
have adopted ${\rm rms}_{\rm NB}=7\, \rm mJy$, which is the mean rms
noise in the follow-up spectra after smoothing these to the \hipass
resolution. The converted error estimates provide good fits to the
measured scatter, indicating that the equations in
Table~\ref{errors.tab} can be used to find reliable errors on \hicat
parameters. 
We note that the errors on the peak and integrated flux include the
uncertainties in the calibration of the flux scale, except for 
errors in the Baars et al. (1977) flux scale.

\subsection{Parameter comparison with Bright Galaxy Catalogue}
 The Bright Galaxy Catalogue (BGC, Koribalski et al. 2004) consists of
the 1000 \hipass galaxies with the highest peak fluxes and is
assembled and parametrized independently from \hicat, but is extracted
from the same data cubes. By comparing the \hicat parameters with
those from the BGC, it can be determined what fraction of the error on
the \hicat parameters is determined by the parametrization procedure
(internal error), and what fraction is caused by noise in the \hipass
data (external error). This comparison is particularly interesting
because generally in the parametrization of a 21-cm emission line
profile a number of choices are made, which could differ between the
persons doing the parametrization. The biggest uncertainty is probably
the fitting and subtracting of the spectral baseline. Structure in the
baseline is caused by ringing associated with strong Galactic \hi\
emission and continuum emission that can produce standing wave
patterns in the telescope structure. For the BGC, the spectral
baselines were fitted with polynomials, of which the order is a free
parameter, whereas \hicat baselines were fitted with Gaussian
smoothing, where the dispersion is a free parameter (see paper
I). Another uncertainty is introduced with the choice of the velocity
extrema of line profiles, between which the flux is integrated.

\begin{figure}
\center{\epsfxsize=10cm \epsfbox[132 167 618 714]{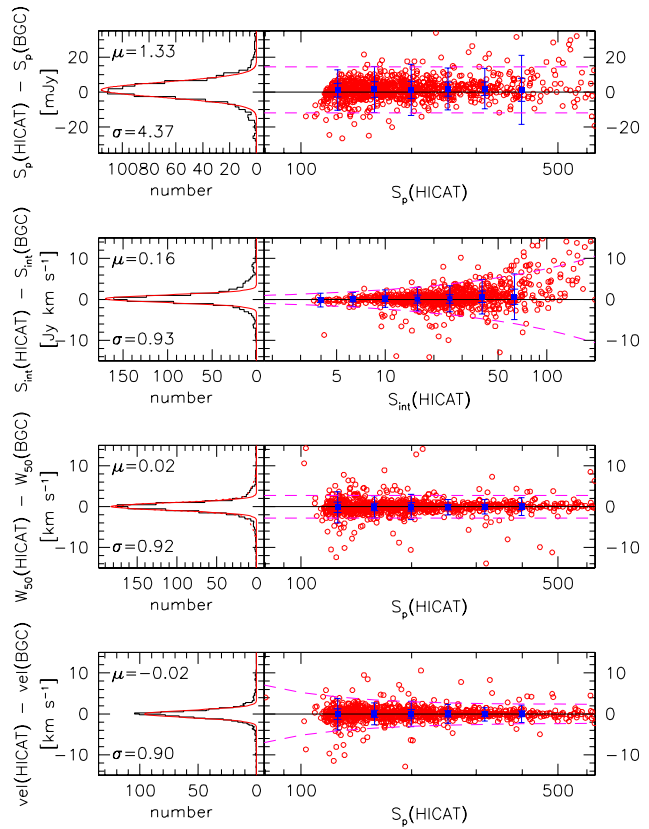}}
\caption{\label{BGCparams.fig} Similar to Fig.~\ref{fakeparams.fig},
  but showing the comparison between parameters from \hicat and
  the \hipass Bright Galaxy Catalogue (Koribalski et al. 2004).}
\end{figure}

In Fig.~\ref{BGCparams.fig} the comparison with the BGC is shown. The
difference between \hicat and BGC parameters is very small.  There are
no systematic trends, except for a slight excess of points with high
values of $\sint({\rm HICAT})-\sint({\rm BGC})$ at large values of
\sint. This excess arises because \hicat and the BGC use different
criteria to define what is an extended source. This leads to more
sources in \hicat being fitted as extended, which generally results in
higher values of \sint. Overall, we find that the parametrization
error contributes only marginally to the total error, with a
contribution of 8 per cent to $\sigma(\speakf)$, 13 per cent to
$\sigma(\sint)$, and 1 per cent to $\sigma(W_{50})$. The contribution
to $\sigma(V_{\rm hel})$ is not uniquely defined because it depends on
\speak, but on average it is 13 per cent.  The rms scatter on the
difference between the BGC and \hicat values of $V_{\rm hel}$ is only
$4.8~\kms$ at the 99 per cent completeness limit and drops to $2~\kms$
for brighter sources.

\section{Summary}
The full catalogue of extragalactic \hipass detections (\hicat) has
now been released to the public (Meyer et al. 2004a, paper I). In the
present paper we have addressed in detail the completeness and
reliability of the survey. We present analytical expressions that can
be used to approximate the completeness and reliability.  We find that
\hicat is 99 per cent complete at a peak flux of 84~mJy and an
integrated flux of $9.4~\jykms$.  The overall reliability is 95 per
cent, but rises to 99 per cent for sources with peak fluxes $>58~\rm
mJy$ or integrated flux $> 8.2~\jykms$.  Expressions are derived for
the uncertainties on the most important \hicat parameters: peak flux,
integrated flux, velocity width, and recessional velocity. The errors
on \hicat parameters are dominated by the noise in the \hipass data,
rather than by the parametrization procedure.

\section*{Acknowledgments}
 The Multibeam system was funded by the Australia Telescope National
Facility (ATNF) and an Australian Research Council grant. The
collaborating institutions are the Universities of Melbourne, Western
Sydney, Sydney, and Cardiff, Mount Stromlo Observatory, Jodrell Bank
Observatory and the ATNF. The Multibeam receiver and correlator was
designed and built by the ATNF with assistance from the Australian
Commonwealth Scientific and Industrial Research Organisation Division
of Telecommunications and Industrial Physics. The low noise amplifiers
were provided by Jodrell Bank Observatory through a grant from the UK
Particle Physics and Astronomy Research Council.  We thank Caroline
Andrzejewski, Alexa Figgess, Alpha Mastrano, Dione Scheltus, and Ivy
Wong for their help with the Parkes narrow-band follow-up
observations.  Finally, we express our sincere gratitude to the staff
of the Parkes Observatory who have provided magnificent observing
support for the survey since the very first \hipass observations in
early 1997.

\label{lastpage}

\end{document}